----------------------------- Start of body part 2

\documentstyle[12pt,epic]{article}

\begin{document}

\begin{titlepage}
\title{Graph Quantum Groups}

\author{C\'esar G\'omez and Germ\'an Sierra
\\ Instituto de Matem\'aticas y F\'{\i}sica
Fundamental \\  CSIC, Serrano 123, E--28006 Spain.}

\date{December 1993}

\maketitle

\begin{abstract}
We define a new mathematical structure (graph quantum group)
which combines the tower
of algebras associated with a graph ${\cal G}$ and the structure of a Hopf
algebra ${\cal A}$. In this structure Ocneanu's string operators
become quantum group intertwiners.
We present some examples of graph quantum groups.

\end{abstract}

\end{titlepage}

\def\a{\alpha}
\def\l{\lambda}
\def\t{\theta}
\def\i{{\rm i}}
\def\s{ {\rm sinh}}
\def\c{ {\rm cosh}}
\def\x{ \xi}
\def\u{$ U_q( \tilde{ {\cal G} })$ }
\def\ua{$ \tilde{U}_q( A_1)$ }
\def\S{ $S-$matrix }
\def\r{ \rho}
\def\A{$ {\cal A}$ }
\def\G{$ {\cal G}$ }
\def\GQG{ $ ({\cal A},{\cal G}, w)$ }

\section{Introduction}
There are two basic mathematical structures underlying the integrability
of two dimensional lattice models, namely quantum groups \cite{DJ}
and Temperley-Lieb-Jones algebras \cite{Pasquier,Harpe} ,
which are associated respectively to vertex models and IRF or
RSOS models \cite{book} . The lattice variables for vertex models
are in correspondence
with irreps of a Hopf algebra \A while for
IRF models they are labelled by the nodes of a given graph \G.
The TLJ algebra is defined using the graph \G
by the Jones fundamental construction \cite{Harpe} . When the graph \G
is defined by means of the decomposition rules of the tensor products of
irreps of \A and the nodes
are associated to these irreps, then a vertex-IRF
map can be defined relating these two types of models \cite{Pasquier} .

It is well known that solvable integrable models can
usually be interpreted as factorizable scattering theories \cite{ZZ}.
In the case of vertex models one finds a scattering of
particles whose internal quantum numbers are governed by irreps
of the Hopf algebra \A \cite{SmR} , which plays the role of a symmetry acting
on asymptotic states. On the other hand the IRF models are more
suitable for a description of scattering of kinks interpolating
between asymptotic vacua configurations. Under this interpretation
the nodes of the graph \G are the vacua and the links are the
kinks \cite{BL} .

The study of $N=2$ integrable massive field theories
has led to the construction of a more general class of
factorizable scattering theories which combine in a non trivial
way the two classes mentioned above .
In these theories the solitons are kinks with extra internal
quantum numbers associated to a Hopf algebra \A and such that
the scattering $S-$matrices depend in general, in a rather complicated
way, on both the Hopf algebra \A and the graph \G
which characterizes the kinks.

This physical problem has motivated us to generalize the definition
of quantum groups in order to include in a non trivial way
the graph \G. We call this new structure graph quantum group
(GQG) \cite{GS}.

\section{Graph Quantum Groups : definition}

In order to define a graph quantum group, \GQG,we will use:

i) A Hopf algebra \A

ii) A connected graph \G

iii) A map $w$ :

\begin{equation}
{\rm Links}({\cal G}) \stackrel{w}{ \longrightarrow}  Rep {\cal A}
\label{1}
\end{equation}

\noindent
which associates to each link of the graph \G an irreducible
finite dimensional representation of \A.

We will reduce our construction to graphs whose incidence matrix
$\Lambda_{a,b}$ ($a,b$ nodes of \G) takes only the values 0 or 1,
i.e there exist at most one link between two nodes.

Given two arbitrary nodes $a,b$ of \G we define the space
$\Omega^{(n)}_{a,b}$ of paths with n links starting at the node
$a$ and finishing at the node $b$:

\begin{equation}
\Omega^{(n)}_{a,b}= \{ \x = (a_0,a_1,\dots,a_{n-1},a_n) \; | \;
a_0=a,a_n=b \;  ; \;   \Lambda_{a_i,a_{i+1}} = 1 \}
\label{2}
\end{equation}

\noindent
We shall use the notation $\x(i)=a_i$ to characterize the value
of the path $\x$ at level $i$.

Using the map $w$ we can associate to each path $\x$ in
$\Omega^{(n)}_{a,b}$ a $n^{th}$ tensor product of irreps of \A:

\begin{eqnarray}
&w: \Omega^{(n)}_{a,b} \rightarrow {\otimes}^n \; Rep \; {\cal A}&
\nonumber \\
&\x \rightarrow w(\x)=  w(a_0,a_1)
\otimes \cdots \otimes w(a_{n-1},a_n) &
\label{3}
\end{eqnarray}

\noindent
We will call a plaquette of \G any couple of paths  $(\gamma^+,\gamma^-)$
of length two starting and finishing at the same points in \G, i.e.
$\gamma^+= (a,d,c)$,
$\gamma^-= (a,b,c)$ such that $\Lambda_{a,b}=\Lambda_{b,c}=\Lambda_{a,d}=
\Lambda_{d,c}=1$. Sometimes we will use the notation
$\left( \begin{array}{@{\,}c@{\,}c@{\,}} a & d \\ b & c
\end{array} \right)$ for representing a plaquette.

{\bf Definition 1.-} For any plaquette $(\gamma^+,\gamma^-)$ of
\G we define the plaquette operators $T_{i}(\gamma^+,\gamma^-)$,
$i=1,..,n-1$, acting on the space of paths
as follows:

\begin{eqnarray}
&\x \rightarrow \x' =
T_i(\gamma^+,\gamma^-) \x & \nonumber \\
&\x'(j) = \prod_{k=0}^2 \delta_{\x(i-1+k),\gamma^-(k)} \;\; \times
\left\{ \begin{array}{ll} \x(j) &  j \neq i \\
\gamma^+(1) & j= i \end{array} \right. &
\label{4}
\end{eqnarray}

\noindent
The vanishing of the Kronecker's delta symbol in eq.(\ref{4})
will mean that the result of the
action of $T_i(\gamma^+,\gamma^-)$ on $\x$ produces
an "empty" path $\phi$.

The plaquette operators can be composed in a path way. They for instance
satisfy:

\begin{eqnarray}
&T_i(\gamma^+,\gamma^-) \; T_i(\eta^+,\eta^-) = \delta_{\gamma^-,
\eta^+} \;\; T_i(\gamma^+,\eta^-)& \nonumber \\
&T_i(\gamma^+,\gamma^-) \; T_j(\eta^+,\eta^-) =
T_j(\eta^+,\eta^-) T_i(\gamma^+,\gamma^-) \;,\; |i-j|\geq 2  &
\label{tt}
\end{eqnarray}

By the map $w$ we can lift the plaquette
operators $T_i(\gamma^+,\gamma^-)$ to
operators $R_i(\gamma^+,\gamma^-)$
acting on tensor products of irreps:

\begin{eqnarray}
&\begin{array}{ccccc}
& \x& \stackrel{w}{ \longrightarrow } & w(\x) & \\
T_i(\gamma^+,\gamma^-)
 & \downarrow & & \downarrow &
R_i(\gamma^+,\gamma^-) \\
& \x' & \stackrel{w}{ \longrightarrow} & w(\x') &
\end{array}&
\label{6}
\end{eqnarray}

So far we have considered \A to be a
non affine Hopf algebra, however
if \A is affine then we can
associate to each link $(a,b)$ an affine irrep:

\begin{equation}
(a,b) \in {\rm Links}({\cal G}) \rightarrow (w(\x),\theta)
\label{8}
\end{equation}

\noindent
where $\theta$ is the affine parameter (see section 3 for
explicit examples).
Interpreting now a plaquette operator $(\gamma^+,\gamma^-)$
as a two particle elastic scattering process, we associate
with it an affine parameter $\theta_{12}$ representing
the relative rapidity, namely for a plaquette
$\left( \begin{array}{@{\,}c@{\,}c@{\,}} a & d \\ b & c
\end{array} \right)$
we use
the following map (\ref{8}):

\begin{eqnarray}
& (a,b) \rightarrow (w(a,b) , \theta_1) &
\nonumber \\
& (b,c) \rightarrow (w(b,c) , \theta_2) &
\nonumber \\
&(a,d) \rightarrow (w(a,d) , \theta_2)&
\nonumber \\
&(d,c) \rightarrow (w(d,c) , \theta_{1})&
\label{10}
\end{eqnarray}

{\bf Definition 2.-} Associated with the map $w$ we define the sets:

\begin{equation}
g^{(n,w)}_{a,b} = \{ \; (\x, w(\x)) \; |\; \x \in \Omega^{(n)}_{a,b} \; \}
\label{12}
\end{equation}

\noindent
and the vector spaces
${\bf C}[g^{(n,w)}_{a,b}]$, which consist of the linear combinations:

\begin{eqnarray}
&v= \sum_{
(a_0,a_1,\dots,a_{n-1}) \in \Omega^{(n)}_{a,b}}
\sum_{m_1,\dots,m_n} \;\;
v^{a_0,\dots,a_n}_{m_1,\dots,m_n} \;
e^{(a_0,a_1)}_{m_1} \otimes \cdots \otimes  e^{(a_{n-1},a_n)}_{m_n} &
\label{13}
\end{eqnarray}

\noindent
where $e^{(a_{i-i},a_i)}_{m_i} (m_i= 1,\dots,{\rm dim}
V_{w(a_{i-1},a_i)}) $ is a basis of the vector space
$V_{w(a_{i-1},a_i)}$.

{\bf Definition 3.-} On ${\bf C}[g^{(n,w)}_{a,b}]$
we define the "Yang-Baxter" operators
${\bf S}_i(\theta) $ as follows;

\begin{eqnarray}
&{\bf S}_i(\theta)\;\; [ \dots e^{(a_{i-1},a_i)}_{m_i} \otimes
e^{(a_i,a_{i+1})}_{m_{i+1}} \dots] =  & \nonumber \\
& \sum_{a'_i,m'_i,m'_{i+1}} \;
S\begin{array}{@{\,}c@{\,}c@{\,}}
m'_{i} & m'_{i+1} \\ m_i & m_{i+1} \end{array}
\left( \begin{array}{@{\,}c@{\,}c@{\,}} a_{i-1} & a'_i \\ a_i & a_{i+1}
\end{array} \right) (\theta)\;
[ \dots e^{(a_{i-1},a'_i)}_{m'_i} \otimes
e^{(a'_i,a_{i+1}) }_{m'_{i+1}} \dots] &
\label{14}
\end{eqnarray}

After these preliminaries we can finally
give the definition of graph quantum group.

{\bf Definition 4.-}  A triplet (\A, \G, $ w$) is a graph quantum
group if there exist plaquette operators $S_i(\gamma^+,\gamma^-)$
such that:

i) {\bf \A-covariance}:

\begin{equation}
S_i(\gamma^+,\gamma^-)(\theta) \;\;
\rho_{w(\gamma^-)} [ \Delta(g) ] =
\rho_{w(\gamma^+)} \;\; [ \Delta(g) ]
S_i(\gamma^+,\gamma^-)(\theta)
\label{15}
\end{equation}

\noindent
for any element $g$ of \A.
By $\rho_{w(\gamma^+)} [ \Delta(g) ]$ we mean the action of
$ \Delta(g)  $ particularized to the representation $w(\gamma^+)$.

ii) {\bf  graph-Yang-Baxter equation (gYB)}: The operators
${\bf S}_i(\theta)$ must satisfy the Yang-Baxter relation \cite{book}:

\begin{eqnarray}
&{\bf S}_i(\theta) \; {\bf S}_{i+1}(\theta+\theta') \;
{\bf S}_i(\theta') =
{\bf S}_{i+1}(\theta') \; {\bf S}_{i}(\theta+\theta') \;
{\bf S}_{i+1}(\theta) & \nonumber \\
&{\bf S}_i(\theta) \; {\bf S}_{j}(\theta') \;=
{\bf S}_{j}(\theta') \; {\bf S}_{i}(\theta) \; \;,\;\; |i-j| \geq 2  &
\label{16}
\end{eqnarray}

Condition i) means that the plaquette operators are intertwiners
of the action of \A. This is equivalent to say that for any plaquette
$(\gamma^+, \gamma^-)$ the representations $w(\gamma^+)$ and
$w(\gamma^-)$ are equivalent.
Using now the matrix representation (\ref{14}) of the
${\bf S}_i$ operators we deduce that the eq.(\ref{16}) is equivalent
to the graph-Yang-Baxter equation (see figure 1 for a graphic representation
of this equation) :

\begin{eqnarray}
\sum_{d,\{ m'\} }
S^{m'_1 m'_2}_{m_1 m_2} \left( \begin{array}{@{\,}c@{\,}c@{\,}}
a_1 & d \\ a_2 & a_3 \end{array}
\right)
(\theta) \;
S^{m'_3 m''_3}_{m'_2 m_3} \left( \begin{array}{@{\,}c@{\,}c@{\,}}
d & a'_3 \\ a_3 & a_4 \end{array} \right)
(\theta+ \theta' ) \;
S^{m''_1 m''_2}_{m'_1 m'_3} \left( \begin{array}{@{\,}c@{\,}c@{\,}}
a_1 & a'_2 \\ d & a'_3 \end{array}
\right)
(\theta')  & &  \nonumber \\
&  &
\label{17} \\
= \sum_{d, \{m'\} }
S^{m'_2 m'_3}_{m_2 m_3} \left( \begin{array}{@{\,}c@{\,}c@{\,}}
a_2 & d \\ a_3 & a_4 \end{array}
\right)
(\theta') \;
S^{m''_1 m'_1}_{m_1 m'_2} \left( \begin{array}{@{\,}c@{\,}c@{\,}}
a_1 & a'_2 \\ a_2 & d \end{array}
\right)
(\theta+ \theta' ) \;
S^{m''_2 m''_3}_{m'_1 m'_3} \left( \begin{array}{@{\,}c@{\,}c@{\,}}
a'_2 & a'_3 \\ d & a_4 \end{array}
\right)
(\theta)  & &
\nonumber
\end{eqnarray}

\begin{figure}
\begin{center}

\unitlength = 1mm
\begin{picture}(140,60)(-30,-30)
\drawline(0,0)(10,17.32)(-10,17.32)(-20,0)(0,0)(10,-17.32)
(-10,-17.32)(-20,0)
\drawline(10,17.32)(20,0)(10,-17.32)
\put(0,0){\circle*{2}}
\put(10,17.32){\circle*{2}}
\put(-10,17.32){\circle*{2}}
\put(-20,0){\circle*{2}}
\put(-10,-17.32){\circle*{2}}
\put(10,-17.32){\circle*{2}}
\put(20,0){\circle*{2}}
\thicklines
\drawline(80,0)(100,0)(90,17.32)(70,17.32)(80,0)(70,-17.32)
(90,-17.32)(100,0)
\drawline(70,17.32)(60,0)(70,-17.32)
\put(80,0){\circle*{2}}
\put(100,0){\circle*{2}}
\put(90,17.32){\circle*{2}}
\put(70,17.32){\circle*{2}}
\put(60,0){\circle*{2}}
\put(70,-17.32){\circle*{2}}
\put(90,-17.32){\circle*{2}}
\put(-24,0){\makebox(0,0){$a_1$}}
\put(56,0){\makebox(0,0){$a_1$}}
\put(-15,-20){\makebox(0,0){$a_2$}}
\put(65,-20){\makebox(0,0){$a_2$}}
\put(-15,20){\makebox(0,0){$a'_2$}}
\put(65,20){\makebox(0,0){$a'_2$}}
\put(-3,2){\makebox(0,0){$d$}}
\put(15,-20){\makebox(0,0){$a_3$}}
\put(95,-20){\makebox(0,0){$a_3$}}
\put(24,0){\makebox(0,0){$a_4$}}
\put(104,0){\makebox(0,0){$a_4$}}
\put(15,20){\makebox(0,0){$a'_3$}}
\put(95,20){\makebox(0,0){$a'_3$}}
\put(82,2){\makebox(0,0){$d$}}
\put(32,0){$=$}
\put(-35,0){\makebox(0,0){$\sum_{d,\{ m' \}}$}}
\put(45,0){\makebox(0,0){$\sum_{d,\{ m' \}}$}}
\dottedline{1.4}(-5,25)(-5,-25)
%\drawline{1.4}(-6.5,23.5)(-5,25)(-3.5,23.5)
%\drawline{1.4}(-19,16)(-20,17.3)(-18.7,17)
%\drawline{1.4}(24.7,9.5)(26,9.2)(25,8)
\dottedline{1.4}(85,25)(85,-25)
\dottedline{1.4}(-20,-17.5)(26,9.2)
\dottedline{1.4}(54,-9.2)(100,17.3)
\dottedline{1.4}(54,9.2)(100,-17.3)
\dottedline{1.4}(-20,17.3)(26,-9.2)
\put(-24,-19.3){\makebox(0,0){$m_1$}}
\put(-5,-28){\makebox(0,0){$m_2$}}
\put(29.3,-11){\makebox(0,0){$m_3$}}
\put(29.3,11){\makebox(0,0){$m''_3$}}
\put(-5,28){\makebox(0,0){$m''_2$}}
\put(-24,19.3){\makebox(0,0){$m''_1$}}
\put(-9,4){\makebox(0,0){$m'_1$}}
\put(-3,-11){\makebox(0,0){$m'_2$}}
\put(6,4){\makebox(0,0){$m'_3$}}
\put(50,-10){\makebox(0,0){$m_1$}}
\put(85,-28){\makebox(0,0){$m_2$}}
\put(102,-20){\makebox(0,0){$m_3$}}
\put(50,10){\makebox(0,0){$m''_1$}}
\put(85,28){\makebox(0,0){$m''_2$}}
\put(102,20){\makebox(0,0){$m''_3$}}
\put(74,4){\makebox(0,0){$m'_1$}}
\put(81,-10){\makebox(0,0){$m'_2$}}
\put(88,4){\makebox(0,0){$m'_3$}}
\end{picture}

\end{center}
\caption[]{graph-Yang-Baxter equation. }
\end{figure}
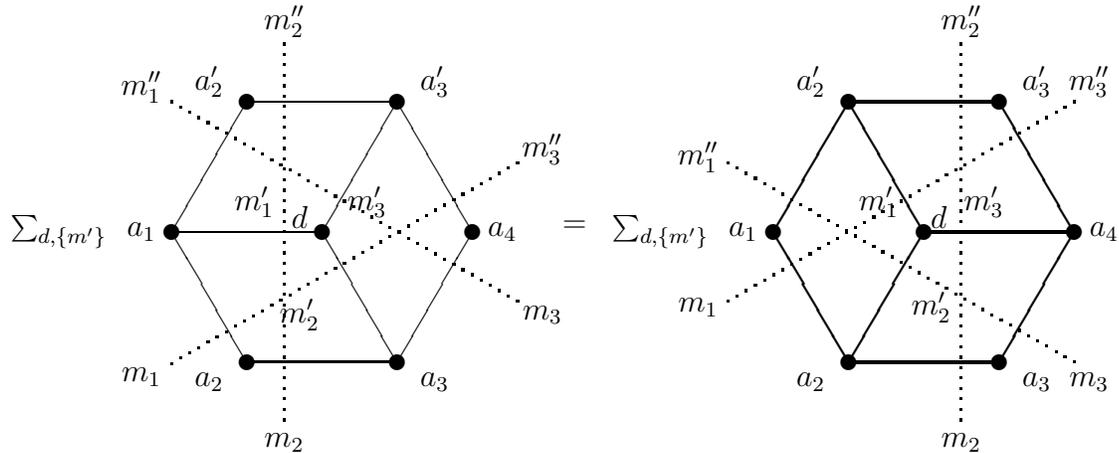

Given two paths $(\eta^+, \eta^-)$ of length $m$ bigger than two
, starting and finishing at the same nodes, i.e.
$\eta^+(0)= \eta^-(0), \eta^+(m)= \eta^-(m)$,
we can construct different tilings of the region
encircled by the path  $\eta^+ \cup \eta^-$
using the plaquettes defined above.
Each tiling give rise to an operator $S(\eta^+,\eta^-)$
( obtained by composing the corresponding plaquette operators).
The gYB equation implies that the operator  $S(\eta^+,\eta^-)$
is tiling independent. By analogy with Ocneanu's path model of
subfactors we shall call $S(\eta^+,\eta^-)$
string operators \cite{Ocne}. These operators
satisfy discrete Migdal-Makeenko
type equations:

\begin{equation}
S(\eta^+,\eta^-) = \sum_{ \tilde{\eta}}  S(\eta^+,\tilde{\eta})
\;\;  S(\tilde{\eta},\eta^-)
\label{MM}
\end{equation}

Using elementary plaquette operators we can also define
discrete path derivatives.
The collection of all string operators of lenght $m$ define an
algebra ${\cal S}_m$. Our construction defines a tower of algebras
$ \cdots \subset {\cal S}_m
\subset {\cal S}_{m+1} \subset \cdots$, which acts
on tensor products of irreps of the Hopf algebra \A. It would
be interesting to know whether this tower defines a subfactor.

{\bf Comments:}

1) The lift by the map $w$ of the plaquette operator,
which according to condition i) of the definition 4
is an intertwiner of the algebra
\A, is not in general an element of the centralizer of \A and therefore
cannot be derived from a universal $R-$matrix.

2) In the definition of GQG's we may restrict ourselves to a certain
family of plaquettes. This restriction should not mean that the reduced
set of plaquettes can be defined by any subgraph of \G. We shall
call this structure restricted graph quantum group (rGQG).

\section{Examples}

{\bf Example 1: N=2 supersymmetric massive theories.}

All integrable $N=2$ supersymmetric massive theories
\cite{Va1,Va2,Wa} define graph quantum
groups \cite{GS} . The lift of the plaquette operator represents the
scattering $S$ matrix for their solitonic spectrum. In many
cases the $S$-matrix operator factorizes into the $R$ matrix of
the $N=2$ Hopf algebra and a solution to the IRF Yang Baxter
equation for a given graph \cite{Fe1,Fe2}.
Next we define the graph quantum
group associated to any integrable $N=2$ massive theory.

i) The Hopf algebra \A is the $N=2$ algebra
generated by the susy charges
$Q^{\pm}$,$\bar{Q}^{\pm}$, the topological central charge $W$
and the fermion number $\cal F$, satisfying the
following (anti)commutation relations:

\begin{eqnarray}
( Q^{\pm})^2=( \bar{Q}^{\pm})^2= &
\{Q^+, \bar{Q}^- \}=  \{Q^-,
 \bar{Q}^+ \}= & 0 \;, \nonumber \\
\{ Q^+,{Q}^- \}= P, & \{ \bar{Q}^+,
 \bar{Q}^- \}= \bar{P} \;, &
\label{19} \\
\{ Q^+, \bar{Q}^+ \} = W,  &
\{ {Q}^-, \bar{Q}^- \}= \bar{W} \;, & \nonumber \\
\left[ {\cal F},Q^{\pm} \right]=\pm Q^{\pm}, &
\left[ {\cal F},\bar{Q}^{\pm} \right]=\mp \bar{Q}^{\pm}
& \nonumber
\end{eqnarray}

\noindent
and comultiplication rules:

\begin{eqnarray}
\Delta Q^{\pm} & = & Q^{\pm} \otimes {\bf 1} +
e^{ \pm {\rm i} \pi \cal{F} } \otimes Q^{\pm} \nonumber \\
\Delta \bar{Q}^{\pm} & = & \bar{Q}^{\pm} \otimes {\bf 1} +
e^{ \mp {\rm i} \pi \cal{F} } \otimes \bar{Q}^{\pm}
\label{20} \\
\Delta W & = & W \otimes {\bf 1} + {\bf 1} \otimes W \nonumber \\
\Delta P & = & P \otimes {\bf 1} + {\bf 1} \otimes P \nonumber
\end{eqnarray}

\noindent
ii) The graph \G is defined by the spectrum as follows
\cite{Ho,Ga} :
The nodes and links of \G represent respectively
the different vacua configurations and the Bogomolnyi solitons.

\noindent
iii) The map $w$ is defined associating to each link $(a,b)$ a two
dimensional irrep of \A,labelled by the casimirs, $m_{a,b}$
which gives the mass, $\Delta_{a,b}$ which is the eigenvalue of
the topological central charge, and the fermion number $f_{a,b}$,
as follows:

\begin{eqnarray}
\pi_{a,b}(\theta)(Q^-) = \left( \begin{array}{cc}
0 & 0 \\ \sqrt{m_{a,b}} e^{\theta/2}  & 0 \end{array} \right),&
\pi_{a,b}(\theta)(Q^+)= \left( \begin{array}{cc}
0 &  \sqrt{m_{a,b}} e^{\theta/2} \\ 0 & 0  \end{array} \right)&
\nonumber \\
\pi_{a,b}(\theta)(\bar{Q}^+) = \left( \begin{array}{cc}
0 & 0 \\ \omega_{a,b}  \sqrt{m_{a,b}}
e^{-\theta/2}  & 0 \end{array} \right),&
\pi_{a,b}(\theta)(\bar{Q}^-)= \left( \begin{array}{cc}
0 & \omega_{a,b}^*  \sqrt{m_{a,b}}
e^{-\theta/2} \\ 0 & 0  \end{array} \right)
& \label{21}   \\
\pi_{a,b}(\theta)({\cal F}) =
\left( \begin{array}{cc} f_{a,b} & 0 \\ 0 & f_{a,b} -1
\end{array} \right)&  & \nonumber
\end{eqnarray}

\noindent
where $\omega_{a,b}=\frac{\Delta_{a,b}}{|\Delta_{a,b}|}$. If the
$N=2$ massive theory admits a lagrangian representation then the
map $w$ and the graph \G can be defined in terms
of a non degenerate Landau-Ginzburg superpotential
$W$ \cite{Va2}. In
this case the nodes of the graph are the critical points of $W$,
the topological charge $\Delta_{a,b}$ is given  by $W_b-W_a$,
where $W_a$ is the critical value of $W$ at the point $a$, and
the fermion numbers are computed from the index theorem formula:

\begin{equation}
{\rm exp}( 2 \pi {\rm i} f_{a, b})
= {\rm phase} \left( \frac{ {\rm det} H(b)}{ {\rm det} H(a)}
\right)
\label{22}
\end{equation}

iv) To define the plaquette operators we consider elastic
processes satisfying equal mass and equal fermion number for
opposite sides.

In table 1 we list the GQG intertwiners, solutions to conditions
i) and ii) of the definition 4, for a reduced set of $N=2$ massive
Landau-Ginzburg theories ( see reference \cite{GS} for details).

\begin{table}
\begin{center}
\begin{tabular}{|c|c|c|}
\hline
Model & Superpotential W & S-matrices  \\ \hline
$A_{k+1}(t_1)$ &$ \frac{x^{k+2}}{k+2} -x$ &
$\begin{array}{c} \mbox{Intertwiners of nilpotent irreps of} \\
U_q(A^{(1)}_1) (q^4=1) \end{array}$
\\ \hline
$A_{k+1}(t_2)$ &$ \frac{x^{k+2}}{k+2} -\frac{x^2}{2}$ &
susy generalization of chiral Potts Boltzmann weights
\\ \hline
$A_{k+1}(t_k) $ & $ \frac{T_{k+2}(x)}{k+2}$ &
$\begin{array}{c} \mbox{Intert. of spin 1/2 irrep of} \\
U_q(A^{(1)}_1) (q^4=1) \end{array}
\times \begin{array}{c} \mbox{Andrews-Baxter-Forrester}
\\ \mbox{ Boltzmann weights} \end{array}$
\\ \hline
$D_{k+3}(\tau)$ &$ \frac{x^{k+2}}{2(k+2)} +\frac{x y^2}{2} -y$&
$ \begin{array}{c} \mbox{ Intertwiners of nilpotent irreps of} \\
\tilde{U}_q( A^{(1)}_1) (q^4=1) \end{array}$
\\ \hline
$D_{k+2}(t_2)$ &$ \frac{x^{k+1}}{2(k+1)} + \frac{x y^2}{2} - x $ &
susy generalization of chiral Potts Boltzmann weights
\\ \hline
$E_6(t_7)$ & $ \frac{x^3}{3} + \frac{y^4}{4} - x y$ &
susy generalization of chiral Potts Boltzmann weights
\\ \hline
$E_8(t_{16})$ & $ \frac{x^3}{3} + \frac{y^5}{5} - x y$ &
susy generalization of chiral Potts Boltzmann weights
\\ \hline
\end{tabular}
\caption{All these examples are perturbations of ADE Landau-Ginzburg
models.The superpotential $T_{k+2}(x)$ of the model $A_{k+1}$
is the Chebishev polynomial $T_n( 2 \; {\rm cos} \t
) = 2 \; {\rm cos} \; n \t$.}
\end{center}
\label{}
\end{table}

{\bf Example 2: Quantum Groups with continuous Spec}

Generically we can equip a given Hopf algebra with a GQG
structure whenever there exist  couples of
equivalent representations of the type:

\begin{equation}
\rho_1 \otimes \rho_2 \simeq \rho_3 \otimes \rho_4
\label{23}
\end{equation}

\noindent
where the irreps $\rho_3$ and $\rho_4$ differ from
$\rho_1$ or $\rho_2$. This
situation can be expected
to occur when the Hopf algebra possesses finite
dimensional irreps parametrized by continuous casimirs, as it is the
case of Hopf algebras at root of unit \cite{KDP}.
To illustrate how this may happen we shall considered the Hopf algebra
\ua \cite{Ji}  with generators $E,F,K,U$. If the deformation parameter
$q$ is a $\ell$ root of unit , i.e. $q^{\ell}=1$, then \ua
has a large center generated, in addition to the usual
cuadratic casimir $C$, by $X=E^{\ell'},\; Y=F^{\ell'},\;
Z=K^{\ell'}$  ($ \ell'= \ell$ if $\ell$ is
odd and $ \ell'= \ell/2 $ if $\ell $ is even )
and the central element $U$.
. These four casimirs form
a central Hopf subalgebra with comultiplications:

\begin{eqnarray}
& \Delta X = X \otimes {\bf 1} + Z \; U^{\ell'} \otimes X &
\nonumber \\
& \Delta Y = Y \otimes Z^{-1} +  U^{-\ell'} \otimes Y &
\label{24} \\
& \Delta Z = Z \otimes Z & \nonumber \\
& \Delta U = U \otimes U &
\nonumber
\end{eqnarray}

The  conditions to have the equivalence shown in equation
(\ref{23}) is that  $X,Y,Z,U$ take the same
values on both representations:

\begin{eqnarray}
&X(\rho_1 \otimes \rho_2) = X(\rho_3 \otimes \rho_4) \Rightarrow
x_1 + z_1 \; u^{\ell'}_1 \;x_2 = x_3 + z_3\; u^{\ell'}_3 \;x_4 & \nonumber \\
&Y(\rho_1 \otimes \rho_2) = Y(\rho_3 \otimes \rho_4) \Rightarrow
y_1\; z_2^{-1} + u^{-\ell'}_1 \;y_2 = y_3 \;z^{-1}_4 +
u^{-\ell'}_3 \;y_4 & \label{25} \\
&Z(\rho_1 \otimes \rho_2) = Z(\rho_3 \otimes \rho_4) \Rightarrow
z_1\; z_2 = z_3 \;z_4 & \nonumber \\
&U(\rho_1 \otimes \rho_2) = U(\rho_3 \otimes \rho_4) \Rightarrow
u_1\; u_2 = u_3\; u_4 &
\nonumber
\end{eqnarray}

For each solution of these equations one is able to find
a ${\ell'}^2 \; \times \; {\ell'}^2$
matrix $R\left( \begin{array}{@{\,}c@{\,}c@{\,}} \rho_3 & \rho_4 \\
\rho_1 & \rho_2 \end{array} \right)$, which is an intertwiner
realizing the equivalence (\ref{23}).

To look for solutions of the graph-Yang-Baxter equation
we need to consider equivalences of the form:

\begin{equation}
\rho_1 \otimes \rho_2 \otimes \rho_3 \simeq
\rho_4 \otimes \rho_5 \otimes \rho_6
\label{26}
\end{equation}

\noindent
which are guaranteed provided:

\begin{equation}
X(\rho_1 \otimes \rho_2 \otimes \rho_3)=
X(\rho_4 \otimes \rho_5 \otimes \rho_6), \;\;{\rm etc}
\label{27}
\end{equation}

To have a gYB one should be able to
factorize eq.(\ref{26}) in the usual Yang-Baxter sequences:

\begin{eqnarray}
& \begin{array}{ccccc}
& & \rho_7 \otimes \rho_8 \otimes \rho_3
\rightarrow \rho_7 \otimes \rho_9 \otimes \rho_6 & & \\
& \nearrow &  &  \searrow & \\
\rho_1 \otimes \rho_2 \otimes \rho_3
& & & & \rho_4 \otimes \rho_5 \otimes \rho_6 \\
& \searrow & & \nearrow &  \\
& &
\rho_1 \otimes \rho_{10} \otimes \rho_{11}
\rightarrow \rho_4 \otimes \rho_{12} \otimes \rho_{11}
 & & \end{array} &
\label{28}
\end{eqnarray}

\noindent
for some intermediate irreps $\rho_7,\dots,\rho_{12}$.
Each step in eq.(\ref{28}) involves a set of eqs. of the type
(\ref{25}), i.e.:

\begin{eqnarray}
&X(\rho_1 \otimes \rho_2) = X(\rho_7 \otimes \rho_8),
X(\rho_2 \otimes \rho_3) = X(\rho_{10} \otimes \rho_{11}) &
\nonumber \\
&X(\rho_8 \otimes \rho_3) = X(\rho_9 \otimes \rho_6),
X(\rho_1 \otimes \rho_{10}) = X(\rho_{4} \otimes \rho_{12}) &
\label{29} \\
&X(\rho_7 \otimes \rho_9) = X(\rho_4 \otimes \rho_5),
X(\rho_{12} \otimes \rho_{11}) = X(\rho_{5} \otimes \rho_{6}) &
\nonumber
\end{eqnarray}

\noindent
and similarly for $Y,Z,U$.
It is easy to see that equations (\ref{29}) with the constraint
(\ref{27}) determine the casimirs $X,Y,Z,U$ of the
six intermediate
irreps but one, which can be choosen at will.
This freedom is the origin of the sum
on IRF labels in the graph-Yang-Baxter equation.

We shall finally make some further comments.
Based on the formal connection between Ocneanu's path model
\cite{Ocne} and Witten's string vertex \cite{Witten},
the construction that we have developed
suggest a possible way to define discrete
strings possessing internal
degrees of freedom governed by quantum groups.
Each string configuration, i.e. a path $\x$ on the Bratteli
diagram of the graph \G, can be associated with a ket
$|\x \rangle $ in the vector space ${\bf C}[\x,w(\x)]$
(see definition 2).
The string operators act now on these string states as prescribed
by equation (\ref{14}).
The interest of this structure is the non-trivial interplay
between the geometry of the strings, characterized by the graph \G,
and the internal degrees of freedom which we associate
with a quantum algebra \A.
Another possible application is the construction of new
integrable lattice models which mix
vertex and IRF degrees of freedom.


\newpage


\begin{thebibliography}{99}

\bibitem{DJ}V.G. Drinfeld,  Proc. ICM-86
(Berkeley) , Vol.1 AMS (1986) 798.

M. Jimbo, Lett.Math.Phys. 10 (1985) 63; 11 (1986) 247.

\bibitem{Pasquier}V. Pasquier, Commun.Math.Phys. 118 (1988) 335.

\bibitem{Harpe}F.M.Goodman, P.de la Harpe and V.F.R.Jones,
"Coxeter-Dynkin diagrams and towers of algebras. MSRI Publications,
Springer-Verlag, 1989.

\bibitem{book}R.J.Baxter, "Exactly solved models in statistical
mechanics", Academic Press (1982).


\bibitem{ZZ}A.B.Zamolodchikov and Al.B.Zamolodchikov,
Ann.Phys. 120 (1979) 235.


\bibitem{SmR}N.Yu.Reshetikhin and  F.Smirnov, Commun.Math.Phys.
131 (1990) 157.

\bibitem{BL}D.Bernard and A. LeClair, Phys.Lett. B247 (1990) 309.

\bibitem{GS}C.Gomez and G.Sierra, CERN-TH.6963/93;UGVA 07/622/93
to appear in Phys.Lett; "On the integrability of $N=2$
supersymmetric massive theories" hepth/9312032.

\bibitem{Ocne}A. Ocneanu, London Math.Soc. Lecture Notes Series
136 (1989).



\bibitem{Va1}C.Vafa and N.P.Warner, Phys.Lett.B218 (1989) 51.

E.Martinec,Phys.Lett.B217 (1989) 431.


\bibitem{Va2}W.Lerche,C.Vafa and N.P.Warner,Nucl.Phys.B324 (1989) 427.

\bibitem{Wa}N. Warner, Proc.1992 Trieste Summer School.

\bibitem{Ho}S.Cecotti and C.Vafa, Nucl.Phys. B367 (1991) 359,
"On the classification of N=2 SUSY theories",
Harvard preprint HUPT-92/A064,SISSA-203/92/EP.

\bibitem{Ga}P.DiFrancesco, F.Lesage and J.B.Zuber,
Saclay preprint SPhT 93/057;hep-th/9306018.


\bibitem{Fe1}P.Fendley,S.D.Mathur,C.Vafa and N.P.Warner, Phys.Lett.
 B243 (1990) 257.

\bibitem{Fe2}P.Fendley and K.Intriligator, Nucl.Phys. B372 (1992) 533;
B380 (1992) 265.

\bibitem{KDP} C.De Concini and V.G.Kac, Progress in Math. 92
(1990) 471.

C.De Concini, V.G.Kac and C.Procesi,"Quantum coadjoint action",
Pisa preprint (1991).

\bibitem{Ji}E.Date, M.Jimbo, M.Miki and T.Miwa:"New R-matrices
associated with cyclic representations of $U_q(A^{(2)}_2)$
", RIMS 706 (1990).


\bibitem{Witten}E.Witten, Nucl.Phys.B268 (1986) 253.

















\end{thebibliography}
\end{document}